\documentclass[fleqn,twoside]{article}
\usepackage{espcrc2}
\usepackage{graphicx}

\begin{document}

\title{Quenched Chiral Log and Light Quark Mass from Overlap Fermions
\thanks{Talk presented by T.\ Draper at Lattice 2002.}
\thanks{This work is supported in part by the U.S. Department of Energy
        under grant numbers DE-FG05-84ER40154 and DE-FG02-02ER45967.}
}

\author{%
Terrence Draper
  \address[UK]{Department of Physics and Astronomy, 
               University of Kentucky, 
               Lexington, KY 40506, USA},
Shao-Jing Dong
  \addressmark[UK],
Ivan Horv\'{a}th
  \addressmark[UK],
Frank Lee
\address{Center for Nuclear Studies, 
         Dept.\ of Physics, 
         George Washington Univ.,
         Washington, DC 20052, USA}
\address{Jefferson Lab, 
         12000 Jefferson Avenue, 
         Newport News, VA 23606, USA},
Keh-Fei Liu
  \addressmark[UK],
Nilmani Mathur
  \addressmark[UK],
and
Jianbo Zhang
\address{CSSM and Dept.\ of Physics and Math.\ Physics,
         Univ.\ of Adelaide, 
         Adelaide, SA 5005, Australia}
       }

\begin{abstract}
We study the quenched chiral behavior of the pion with mass as low as $\approx
180$ MeV. The calculation is done on a quenched lattice of size $16^3\times 28$
and $a = 0.2\,{\rm fm}$ with 80 configurations using overlap fermions and an
improved gauge action.  Using an improved constrained curve fitting technique,
we find that the ground state pseudoscalar mass versus bare quark mass behavior
is well controlled with small statistical errors; this permits a reliable fit
of the quenched chiral log effects, a determination of the chiral log parameter
($\delta = 0.26(3)$), and an estimate of the renormalized mass of the light
quark ($m^{\overline{MS}}(\mu=2\,{\rm GeV}) = 3.7(3)\,{\rm MeV}$).
\end{abstract}

\maketitle


\section{Simulation Details}

Using a $\beta=2.264$ renormalization-group-improved Iwasaki~\cite{Iwa85} gauge
action, we study the chiral properties of hadrons on a $16^3\times 28$ lattice
with the overlap fermion~\cite{Neu98,Nar95} and massive overlap
operator~\cite{Ale00,Cap01,Her01}
\begin{eqnarray*}
	D(m_0) 
	& = &
	(\rho + \frac{m_0a}{2}) + (\rho - \frac{m_0a}{2} ) \gamma_5 \epsilon (H)
\end{eqnarray*}
where $\epsilon (H) = H /\sqrt{H^2}$, $H = \gamma_5 D_w$, and $D_w$ is the
usual Wilson fermion operator, except with a negative mass parameter $-\rho =
1/2\kappa -4$ in which $\kappa_c < \kappa < 0.25$; we take $\kappa = 0.19$ in
our calculation which corresponds to $\rho = 1.368$.

We use the optimal partial fraction expansion with a 14th-order Zolotarev
approximation of the matrix sign function~\cite{Esh02}; the sign function
approximated to better than $3$ parts in $10^{10}$.  As the conjugate-gradient
inverter accommodates multi-mass~\cite{Fro95} we obtain the quark propagator at
$26$ masses including $18$ masses at or below the strange quark mass with less
than 10\% overhead.

\vfill


\section{Pion Decay Constant and $Z_{A}$}

The renormalized pion decay constant is
\begin{eqnarray*}
	f_{\pi}^{(R)} \sqrt{2}
	& = & 
	Z_{A} f_{\pi}^{(U)} \sqrt{2}  \\
	& = &
	\frac{f_{P}^{(U)}}{(m_{\pi}^2/2m_0)}
	=
	\frac{\sqrt{2m_{\pi}A_{PP}}}{(m_{\pi}^2/2m_0)}
\end{eqnarray*}
where $m_0$ is the unrenormalized quark mass,
$f_{P}^{(U)}=\sqrt{2m_{\pi}A_{PP}}$ is the unrenormalized pseudoscalar decay
constant, and $A_{PP}$ is the amplitude of the two-point local-local
correlator, $G_{PP}(t) \longrightarrow A_{PP} ( e^{-m_{\pi}t} +
e^{-m_{\pi}(T-t)})$.  Since $f_{\pi}^{(R)}$ is free of quenched chiral logs,
and since we determine it quite precisely, we use $f_{\pi}^{(R)}$ to set the
scale and obtain $a^{-1}=0.978(5)\,{\rm GeV}$.

As a bonus, we obtain the axial-vector renormalization constant, $Z_A$, via the
axial Ward identity $Z_{A} \partial_{\mu} A_{\mu} = 2 Z_{m} m_{0} Z_{P}P$ and
the relations $Z_{m}^{-1} = Z_{P} = Z_{S}$ protected by the chiral symmetry of
the overlap fermion.  Thus $Z_{A}= \frac{2m_{0}}{m_{\pi}}
\sqrt{\frac{A_{PP}}{A_{A_{4}P}}}$ which is plotted in Fig.~1.
\begin{figure}[t]
  \begin{center}
  \includegraphics[angle=0,width=\hsize]{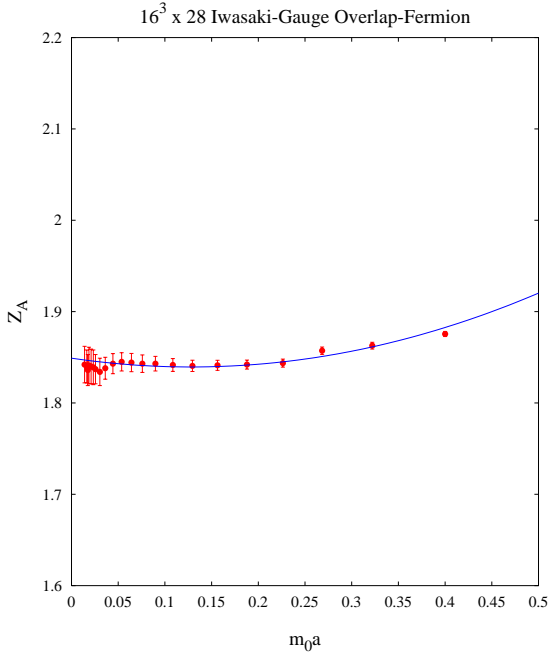}
  \vspace{-1.5cm}
  \caption {$Z_A$ {\it vs}.\ bare mass, $m_{0}a$.}
  \vspace{-1.0cm}
  \end{center}
\end{figure}

A covariant polynomial fit yields
\begin{eqnarray*}
	Z_{A} 
& = & 
	1.85(1) - 0.60(12) m_{0}\Lambda_{\rm QCD}a^2 \\ 
&   &   + 0.59(4) m_{0}^2a^2 \quad ; \Lambda_{\rm QCD}a=0.25
\end{eqnarray*}
so ${\cal O}(m_{0}\Lambda_{\rm QCD}a^2)$ and ${\cal O}(m_{0}^2a^2)$ terms are
small.

\vfill


\section{Pion Mass and Chiral Logs}

\begin{figure}[t]
  \begin{center}
  \includegraphics[angle=0,width=\hsize]{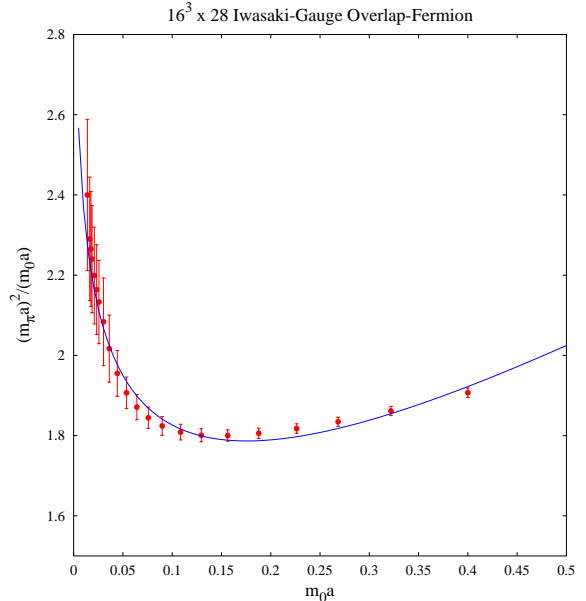}
  \vspace{-1.5cm}
  \caption {$(m_{\pi}a)^2 / m_{0}a$ {\it vs}.\ bare quark mass, $m_{0}a$.}
  \vspace{-1.0cm}
  \end{center}
\end{figure}

Our plot of $(m_{\pi}a)^2 / m_{0}a$ in Fig.~2, with $18$ data points at or
below the strange quark mass, dramatically reveals the presence of the quenched
chiral log.
We fit $m_{\pi}^2 a^2$ to the form~\cite{Ber92,CPP00}
\begin{eqnarray*}
	m_{\pi}^2 a^{2} 
	& = & 
	A m_{0}a \left\{1 - \delta 
	\left[\ln\left(\frac{Am_{0}a}{\Lambda_{\chi}^{2}a^{2}}\right)+1\right]\right\}
\end{eqnarray*}
using a single multi-parameter (weakly) constrained fit (the constraint lifts
the degeneracy of the three-parameter fit).  We obtain simultaneously (the
preliminary results) $\delta = 0.26(3)$ and $\Lambda_{\chi}a=1.1(1)$, which
{\it are quite stable\/} against changing the window, [$m_{\pi}^{\rm
min}$,$m_{\pi}^{\rm max}$], of masses in the fit provided (a) $m_{\pi}^{\rm
min}<250\,{\rm MeV}$; thereafter, $\delta$ monotonically decreases rapidly with
increasing $m_{\pi}^{\rm min}$ (e.g.\ $0.14$ for $300\,{\rm MeV}$, $0.06$ for
$400\,{\rm MeV}$), and (b) $m_{\pi}^{\rm max}<350\,{\rm MeV}$; adding a term $B
m_{0}^{2}a^{2}$ to the fit formula increases the region of stability to
$m_{\pi}^{\rm max}>700\,{\rm MeV}$.

The pseudoscalar decay constant $f_{P}$ has similar behavior and fit results to
$(m_{\pi}a)^2 / m_{0}a$.  (This is automatic for overlap fermions since
$Z_m^{-1}=Z_P=Z_S$ implies $f_{P}=f_{\pi}^{(R)} \sqrt{2} m_{\pi}^2/2m_0$.)

Our value of $\delta$ is significantly higher than other recent
values~\cite{CPP00,Bar00,QCD00,MIL01,Chi02} most of which are well below the
prediction of $\delta=0.183$ from the quenched $\eta'$ loop (using the
Witten-Veneziano model of the $\eta'$ mass) in chiral perturbation theory;
however, we have exact chiral symmetry and, to date, use the largest spatial
volume and probe the lightest pion masses.  Indeed, the effect of the chiral
log makes the data points rise as the pion mass drops below $m_{\pi} \approx
400$--$450\,{\rm MeV}$. Notice that if one restricts the data to pion masses in
the range of $350$--$400\,{\rm MeV}$ to $700\,{\rm MeV}$, then the curve can be
well approximated by a constant within errors.  Equivalently, a plot of
$m_{\pi}^2$ versus $m_{0}$ over this range would show no deviation from
linearity, as is also seen in~\cite{Her01,Giu01}.  The effects of chiral logs
are seen only at lower pion mass; one is unlikely to see the effect at all
without data below $m_{\pi} \approx 400\,{\rm MeV}$.  With adequate spatial
volume, the extraction of $\delta$ would give a low value without data below
$m_{\pi} \approx 250\,{\rm MeV}$.  Our spatial lattice size is $L = 3.2\,{\rm
fm}$ so finite-volume effects are expected to be small; our lightest
(unrenormalized) quark mass of $\approx 14\,{\rm MeV}$ corresponds to a pion of
about $180\,{\rm MeV}$.  With a lattice scale of $a=0.202(1)\,{\rm fm}$ (set
from $f_{\pi}$) then $L=3.2\,{\rm fm}$ and $(Lm_{\pi})_{\rm lightest}\approx
2.9$.

\vfill


\section{Quark Masses}

To renormalize our estimates of the quark mass, we follow~\cite{Her01} to
determine the renormalization constants $Z_{m}^{-1}=Z_S=Z_P$ and thus $m^{RGI}
= Z_{m}(g_0)m_{0}(g_{0})$: The value of the bare quark mass $m_{0}a$ that
reproduces a given fiducial pseudoscalar mass (through $x_{\rm
ref}=(r_{0}m_{\pi})^2$, with $r_{0}/a=2.885(18)$ for our lattice) is obtained
by interpolation of our data for the pion mass.  From this we obtain
$Z_{m}^{-1}=Z_{S}$ at three values of quark mass (the lowest two quark masses
are roughly half-strange and strange).  The operators and action are ${\cal
O}(a)$ improved.  For $Z_A$ we find small ${\cal O}(m_{0}\Lambda_{\rm QCD}a^2)$
and ${\cal O}(m_{0}^2a^2)$ dependence as well; $Z_{A}$ is well approximated (to
$\approx 1\%$) by a constant up through the strange mass region.  Thus we fit
$Z_{m}^{-1}=Z_{S}$ to a constant and obtain $Z_{S}=1.61(7)$ as a first
approximation.

\begin{figure}[t]
  \begin{center}
  \includegraphics[angle=0,width=\hsize]{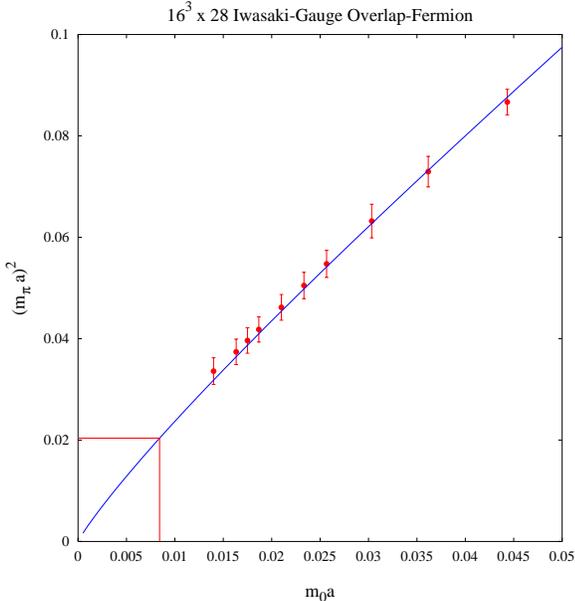}
  \vspace{-1.5cm}
  \caption {$(m_{\pi}a)^2$ {\it vs}.\ pion mass, $m_{\pi}a$.  The solid curve
            is the the chiral log fit. The solid lines indicate the intercepts
            at the physical values.}
  \vspace{-1.0cm}
  \end{center}
\end{figure}

Our lightest pion mass is $\approx 180\,{\rm MeV}$ so we needn't extrapolate
far to obtain $m_{0}a$ at the physical pion mass (Fig.~3).  Furthermore, exact
chiral symmetry (zero residual mass) and the form of the fit function as
determined by quenched ChPT provide an end point at zero mass.

Using $m^{RGI}= m_{0}/Z_{S}$ and, from the 4-loop calculation with
$N_f=0$~\cite{ALP00}, $Z_{S}^{\overline{MS}}=Z_{S}/0.72076$, we obtain (very
preliminary results) for $(m_{u}+m_{d})/2$: $m^{RGI}=5.1(4)\,{\rm MeV}$ and
\begin{eqnarray*}
	m^{\overline{MS}}(\mu=2\,{\rm GeV})
	& = & 
	3.7(3)\,{\rm MeV} 	
\end{eqnarray*}

Last year's result~\cite{Don02} from a $20^4$ lattice with a L\"{u}scher-Weisz
gauge action was $4.5(3)\,{\rm MeV}$. (The result quoted in the proceedings was
an average of this number with an attempt to subtract out the effects of the
chiral log.)  A recent estimate~\cite{Lub01} of the world average is
$4.5(6){\rm MeV}$.


\end{document}